\documentclass[aps,groupaddress,floats,preprint,showpacs]{revtex4}
\usepackage{amsfonts}
\usepackage{amssymb}
\usepackage{bm}
\usepackage[final]{graphicx}

\begin{document}
\bibliographystyle{revtex}


\newcommand{\alps}{\ensuremath{\alpha_s}}
\newcommand{\qbar}{\bar{q}}
\newcommand{\beq}{\begin{equation}}
\newcommand{\eeq}{\end{equation}}
\newcommand{\beqa}{\begin{eqnarray}}
\newcommand{\eeqa}{\end{eqnarray}}
\newcommand{\mq}{m_Q}
\newcommand{\mn}{m_N}
\newcommand{\bb}{\langle}
\newcommand{\kb}{\rangle}
\newcommand{\st}{\ensuremath{\sqrt{\sigma}}}
\newcommand{\rvec}{\mathbf{r}}
\newcommand{\bvec}[1]{\ensuremath{\mathbf{#1}}}
\newcommand{\bra}[1]{\ensuremath{\bb#1|}}
\newcommand{\ket}[1]{\ensuremath{|#1\kb}}
\newcommand{\gft}{\ensuremath{\gamma_{FT}}}
\newcommand{\bfsig}{\mbox{\boldmath{$\sigma$}}}
\newcommand{\bfnab}{\mbox{\boldmath{$\nabla$}}}
\newcommand{\bftau}{\mbox{\boldmath{$\tau$}}}
\newcommand{\spup}{\uparrow}
\newcommand{\spd}{\downarrow}
\newcommand{\hbarom}{\frac{\hbar^2}{m_Q}}
\newcommand{\vnn}{\ensuremath{\hat{v}_{NN}}}
\newcommand{\argonne}{\ensuremath{v_{18}}}
\newcommand{\lqcd}{\ensuremath{\mathcal{L}_{QCD}}}
\newcommand{\lgf}{\ensuremath{\mathcal{L}_g}}
\newcommand{\lqm}{\ensuremath{\mathcal{L}_q}}
\newcommand{\lqg}{\ensuremath{\mathcal{L}_{qg}}}
\newcommand{\nn}{\ensuremath{NN}}
\newcommand{\hpnd}{\ensuremath{H_{\pi N\Delta}}}
\newcommand{\hpqq}{\ensuremath{H_{\pi qq}}}
\newcommand{\fpnn}{\ensuremath{f_{\pi NN}}}
\newcommand{\fpnd}{\ensuremath{f_{\pi N\Delta}}}
\newcommand{\fpqq}{\ensuremath{f_{\pi qq}}}
\newcommand{\ylm}{\ensuremath{Y_\ell^m}}
\newcommand{\ylmc}{\ensuremath{Y_\ell^{m*}}}
\newcommand{\qbh}{\hat{\bvec{q}}}
\newcommand{\xbh}{\hat{\bvec{X}}}
\newcommand{\dt}{\Delta\tau}
\newcommand{\qmag}{|\bvec{q}|}
\newcommand{\oas}{\ensuremath{\mathcal{O}(\alpha_s)}}
\newcommand{\vtxb}{\ensuremath{\Lambda_\mu(p',p)}}
\newcommand{\vtxp}{\ensuremath{\Lambda^\mu(p',p)}}

\title
{Scaling of space and timelike response of confined relativistic particles}

\author{Mark W.\ Paris}
\email[]{paris@uiuc.edu}
\author{Vijay R.\ Pandharipande}
\email[]{vrp@uiuc.edu}
\affiliation{Department of Physics,
University of Illinois at Urbana-Champaign,
1110 West Green Street, Urbana, Illinois 61801}

\date{\today}

\begin{abstract}
\medskip
The response of a relativistic particle bound in a linear confining
well is calculated as a function of the momentum and energy transfer,
$\bvec{q}$, $\nu$. At large values of $\qmag$ the response exhibits
scaling in the variable $\tilde{y}=\nu-\qmag$, which is proportional
to the Nachtmann variable, $\xi$. The approach to scaling is
studied at smaller values of $\qmag$.  Scaling occurs at
$\nu \sim \qmag$ at relatively small $\qmag$,  and its validity  
extends over the entire $\xi$ range as $\qmag$ increases; this
behavior is observed in electron-proton scattering.
About $10\%$ of the response at large $\qmag$ is in the
timelike region where $\nu>\qmag$, and it is necessary to
include it to fulfill the particle number sum rule.
The Gross-Llewellyn Smith and Gottfried sum rules are
discussed in the context of these results. 
\end{abstract}
\maketitle

Deep inelastic scattering (DIS) of leptons by hadrons is generally
discussed in the framework of the naive parton model and the
QCD-improved parton model using the operator product expansion.\cite{ESW}
This approach has been very successful in determining the
evolution of the structure functions as a function of the square
of the four-momentum transferred to the hadron.\cite{AP77}
In the leading order of the model the hadron is approximated by a 
collection of noninteracting quarks and gluons. 
The struck quark is assumed to be on the mass-shell
both before and after its interaction with the electron. Based on  
this assumption, the leading order response is predicted to be in 
the spacelike region for which the energy transfer $\nu$ is less
than the magnitude of momentum transfer, $\qmag$, as a consequence
of the inequality,
\beq
\label{eqn:nupwia}
\nu = \sqrt{|\bvec{k}+\bvec{q}|^2+m_q^2} - \sqrt{|\bvec{k}|^2+m_q^2}
\leqslant \qmag.
\eeq
Here $\bvec{k}$ and $m_q$ are the momentum and mass of the struck
quark, respectively.  The predicted response is discontinuous at 
the boundary $|\bvec{q}|=\nu$ between space and timelike regions. 

\begin{figure}[t]
\includegraphics[ width=300pt, bb=0 0 640 640,
                  keepaspectratio, clip, angle=-90 ]{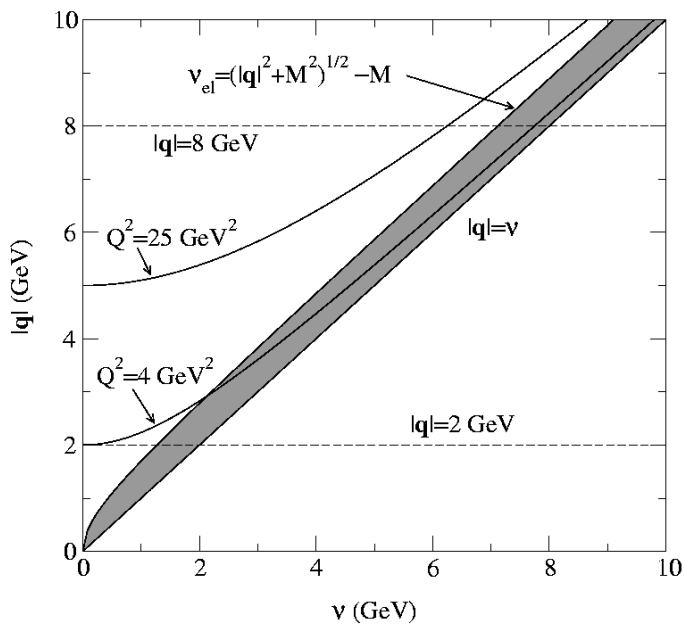}
\nopagebreak
\caption{\label{fig:qvn} The $\qmag$-$\nu$ plane. The spacelike region
is above the $\qmag=\nu$ line and timelike is below. Lines of constant
$Q^2 > 0$ are parabolas which lie entirely in the spacelike region
and approach $\qmag=\nu$ as $\nu\rightarrow\infty$. The observed
($Q^2>0$) response of the proton lies in the shaded area.}
\end{figure}

The conventional variables of the parton model,
$Q^2 = |\bvec{q}|^2 - \nu^2 $ and
the Bjorken $x=Q^2/2M\nu$, used to describe the DIS structure
functions of a hadron of mass $M$, are confined to the spacelike
region of the $\qmag$-$\nu$ plane for positive values of $Q^2$ 
accessible in lepton scattering experiments, as shown in
Fig.\ \ref{fig:qvn}. Therefore we study the response,
$R(\bvec{q},\nu)$ as a function of $\nu$ and $\qmag$ in the rest
frame of the system \cite{BPS00}, as is common practice
in the many-body theory (MBT). Lines of constant $\qmag$
in Fig.\ \ref{fig:qvn} cross
the photon line ($\nu=\qmag$) and go into the timelike region.
The observed ($Q^2>0$) DIS response is limited to a narrow region 
in the $\qmag$-$\nu$ plane illustrated in Fig.\ \ref{fig:qvn}. 
It is bounded by the elastic limit, $\nu_{el}=\sqrt{\qmag^2+M^2}-M$
on one side, and by the photon line on the other.  
In the limit of large $\qmag$ the width of the observed response 
at fixed $\qmag$ is $M$.
Lines of constant $Q^2$ intersect the elastic limit curve at
$x=1$ and approach the photon line at small $x$.

For a hypothetical scalar probe, the response is given by:
\beq
\label{eqn:rqn}
R(\bvec{q},\nu) = \sum_I |\bra{I} \sum_j e^{i \bvec{q}\cdot\rvec_j}
\ket{0}|^2 \delta(E_I - E_0 - \nu)
\eeq
where $\sum_j$ is over all the particles and the $\sum_I$
over all energy eigenstates. It is viewed
as the distribution of the strength of the state 
$\sum_j e^{i\bvec{q}\cdot\rvec_j}
\ket{0}$ over the energy eigenstates of the system having 
momentum $\bvec{q}$. It is
not necessarily zero in the timelike, $\nu > |\bvec{q}|$ region.

The natural scaling variable in the MBT approach to 
DIS \cite{BPS00} is $\tilde{y}=\nu-|\bvec{q}|$. 
At large $|\bvec{q}|$ the response is expected to depend only on 
$\tilde{y}$, and not on $\bvec{q}$ and $\nu$ independently. This
variable is equivalent to the Nachtmann variable $\xi$ since
\cite{ON,Jaffe85}
\beq
\label{eqn:xi}
\xi = \frac{1}{M}(|\bvec{q}|-\nu) = -\frac{1}{M} \tilde{y}.
\eeq
In the limit of large $Q^2$ the 
$\xi = x$, thus $\tilde{y}$ scaling includes Bjorken scaling.  
However, both $\tilde{y}$ and $\xi$ span both spacelike 
and timelike regions at fixed $|\bvec{q}|$ unlike $x$ at fixed $Q^2$. 

The particle number sum rule in MBT is obtained by integrating the 
response at large $|\bvec{q}|$ over all $\nu > 0$:
\beqa
\int_0^{\infty}R(\bvec{q},\nu)d\nu &=& \sum_I \langle 0|\sum_i 
e^{-i\bvec{q}\cdot\bvec{r}_i} |I\rangle \langle I | \sum_j 
e^{i\bvec{q}\cdot\bvec{r}_j }|0\rangle   \nonumber \\
&=& \sum_{i,j} \langle 0|e^{i\bvec{q}\cdot (\bvec{r}_j-\bvec{r}_i)}|0\rangle.
\eeqa
When $\bvec{q}$ is large only the $i=j$ terms in the above
sum contribute, and therefore the integral gives the number
of particles in the system. In contrast the sums of the response
in the parton model are obtained by integrating the response
over $\xi > 0$ at fixed $Q^2$.  These sums will fulfill the
particle number sum rule only if the
response in the timelike region is zero. As mentioned earlier,
the response of a collection of noninteracting particles lies in
the spacelike region. Interaction effects, however, can shift a
part of the strength to the timelike region. Evidence for shifts
caused by interactions is discussed in Ref.\ \cite{BPS00}.

\begin{figure}[t]
\includegraphics[ width=240pt, keepaspectratio, clip, angle=-90 ]{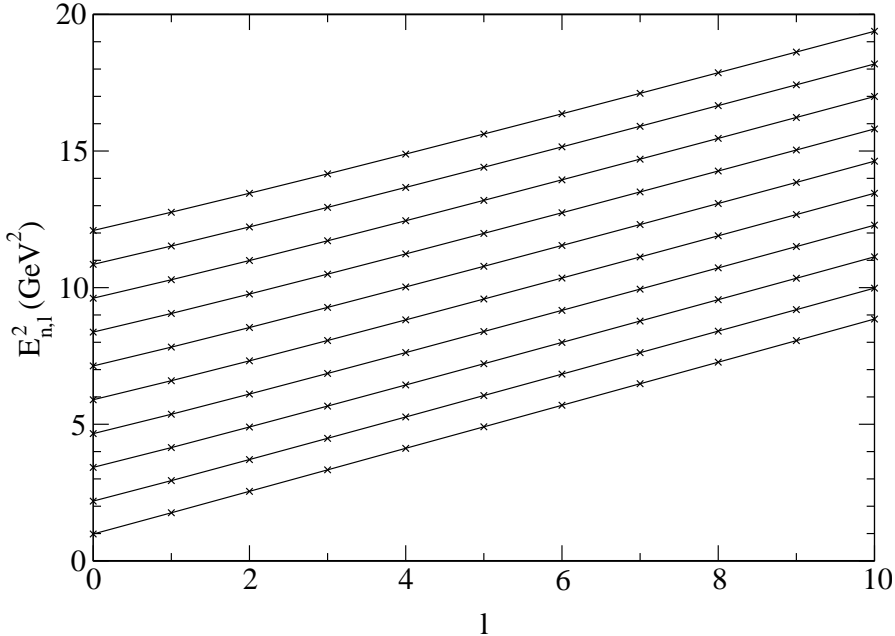}
\nopagebreak
\caption{\label{fig:spec}
$E^2_{n,\ell}$ in GeV$^2$ plotted against angular momentum $\ell$
for the first ten values of the radial quantum number $n$.
The lines are linear fits to the calculated values.}
\end{figure}

We have studied the exact response of a simple ``toy'' model
which contains the basic features of relativity and confinement
to obtain further insights on the possible response in the
timelike region and it's effects on the sum rules. In this model
we assume that the response of the hadron is due to a single
light valence quark confined within the hadron by its
interaction with an infinitely massive color charge. We model
this interaction by a linear flux-tube potential, and use the
single particle Hamiltonian,
\beq
\label{eqn:h1}
H = \sqrt{|{\bvec{p}}|^2+m_q^2} + \st \ r
\eeq
containing the relativistic kinetic energy operator. 
In the limit $m_q=0$ used here, the $H$ can be cast in the form:
\beq
H=\sigma^{1/4} \left( \sqrt{|\bvec{p^{\prime}}|^2} + r^{\prime} \right), 
\eeq
where $\bvec{p}'= \bvec{p}/\sigma^{1/4}$,
and $\bvec{r}'=\sigma^{1/4} \bvec{r}$ are dimensionless. 
The response $R(\qmag , \nu)$ of the model then depends only on the 
dimensionless variables $|\bvec{q}'|=\qmag /\sigma^{1/4}$ and 
$\nu'=\nu/\sigma^{1/4}$.  The main conclusions of this work are 
independent of the assumed value of $\sigma$; however, we show results 
in familiar units using the typical value $\sqrt{\sigma}=1$ GeV/fm. 

The model may be viewed as that of a meson with a heavy antiquark
or that of a baryon with a heavy diquark.  It is obviously 
too simple to address the observed response of hadrons.  For 
example, it omits the sea quarks and radiative gluon effects contained
in the DGLAP equations \cite{AP77,ESW} to describe scaling violations. 
Nevertheless its exact solutions are interesting and useful to study 
scaling, the approach to scaling, and the contribution of the timelike
region to sum rules.  A similar model has been considered by Isgur 
{\em et al.} \cite{Isgur01}. 

\begin{figure}[t]
\includegraphics[ width=260pt, keepaspectratio, clip, angle=-90 ]{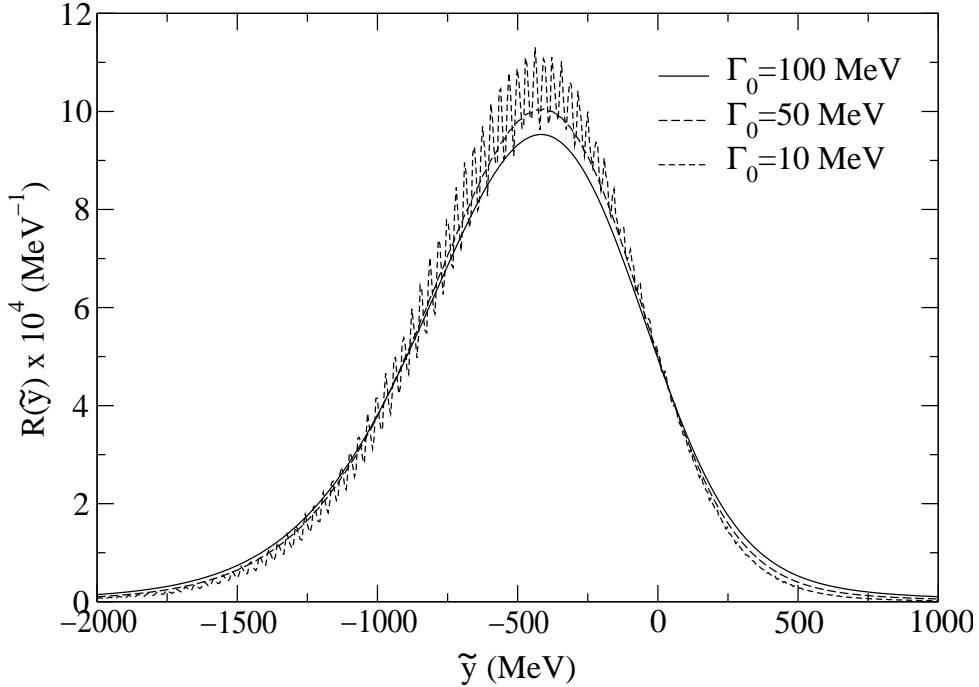}
\nopagebreak
\caption{\label{fig:rvg}
The response for $\qmag=10$ GeV versus the variable,
$\tilde{y}=\nu-\qmag$, for various $\Gamma_0$.}
\end{figure}

The eigenstates of this Hamiltonian have the usual quantum
numbers $n$, $\ell$, and $m$. They are expanded in the basis of 
$j_\ell(k^\ell_\alpha r) Y_\ell^m(\hat{\rvec})$. For each value of
$\ell\leqslant 100$ we consider 200 $(\alpha=1,\ldots,200)$
basis states each having $j_\ell(k^\ell_\alpha R)=0$, for
$R=10$ fm, and 300 basis states for $R=15$ fm. The Hamiltonian
is diagonalized within a sphere of radius $R$. The ground and
relevant excited states have rather small radii, and are not
influenced by this boundary condition. For example, the energies
of the lowest 100 eigenstates of each $\ell$ and the calculated responses
at $\qmag\leqslant 10$ GeV are essentially the same for $R=10$ and
15 fm. Figure \ref{fig:spec} shows a plot of the square of
the energy eigenvalue for a given $n$ versus $\ell$. The calculated
energies lie on linear Regge trajectories
having slopes within $10 \%$ of the classical estimate 
$4 \sqrt{\sigma}$.

The response is calculated including all states
with $n\leqslant 100$ and $\ell\leqslant 100$ in Eq.(\ref{eqn:rqn}).
It is a sequence of $\delta$ functions at $\nu=E_{n,\ell}-E_0$. We verify
that the full strength of the integrated response, 
\beq 
\label{eqn:sumr}
\int_0^{\infty} R(\bvec{q},\nu) d\nu = 1, 
\eeq
is obtained in this truncated basis
for all values of the momentum transfer considered in this work
with $<$ 0.02 \% error.
In order to obtain a smooth response
we assume decay widths for all the excited states  
dependent on the excitation energy $\nu$: 
\beqa
\label{eqn:gam}
\Gamma(\nu \geqslant E_t)&=&\Gamma_0 \left(1-e^{-(\nu-E_t)/E_s}\right). 
\eeqa
Here $E_t$ is the threshold excitation energy for meson
emission below which the 
width is zero, and $E_s$ parameterizes
the approach to a constant width $\Gamma_0$ at $\nu > E_t$.
We use $E_t=E_s=100$ MeV and various values of $\Gamma_0$ for illustration. 
Only the response at small values of $\qmag$ is sensitive to $E_t$ and 
$E_s$.  The response including decay widths is given by: 
\beq
\label{eqn:smear}
R(\bvec{q},\nu) = \sum_I|\langle I| e^{i \bvec{q} \cdot \bvec{r}}
|0 \rangle|^2 \left(\frac{\Gamma(\nu)}{2\pi}\right) \frac{1}
{(E_I-E_0-\nu)^2 + \Gamma^2(\nu)/4} \ . 
\eeq 
The dependence of the response at $\qmag=10$ GeV on $\Gamma_0$ 
is shown in Fig.\ \ref{fig:rvg}.  Since this response has $\nu\gg E_t$ 
and $E_s$, it depends only on $\Gamma_0$. 

\begin{figure}[t]
\includegraphics[ width=240pt, keepaspectratio, clip, angle=-90 ]{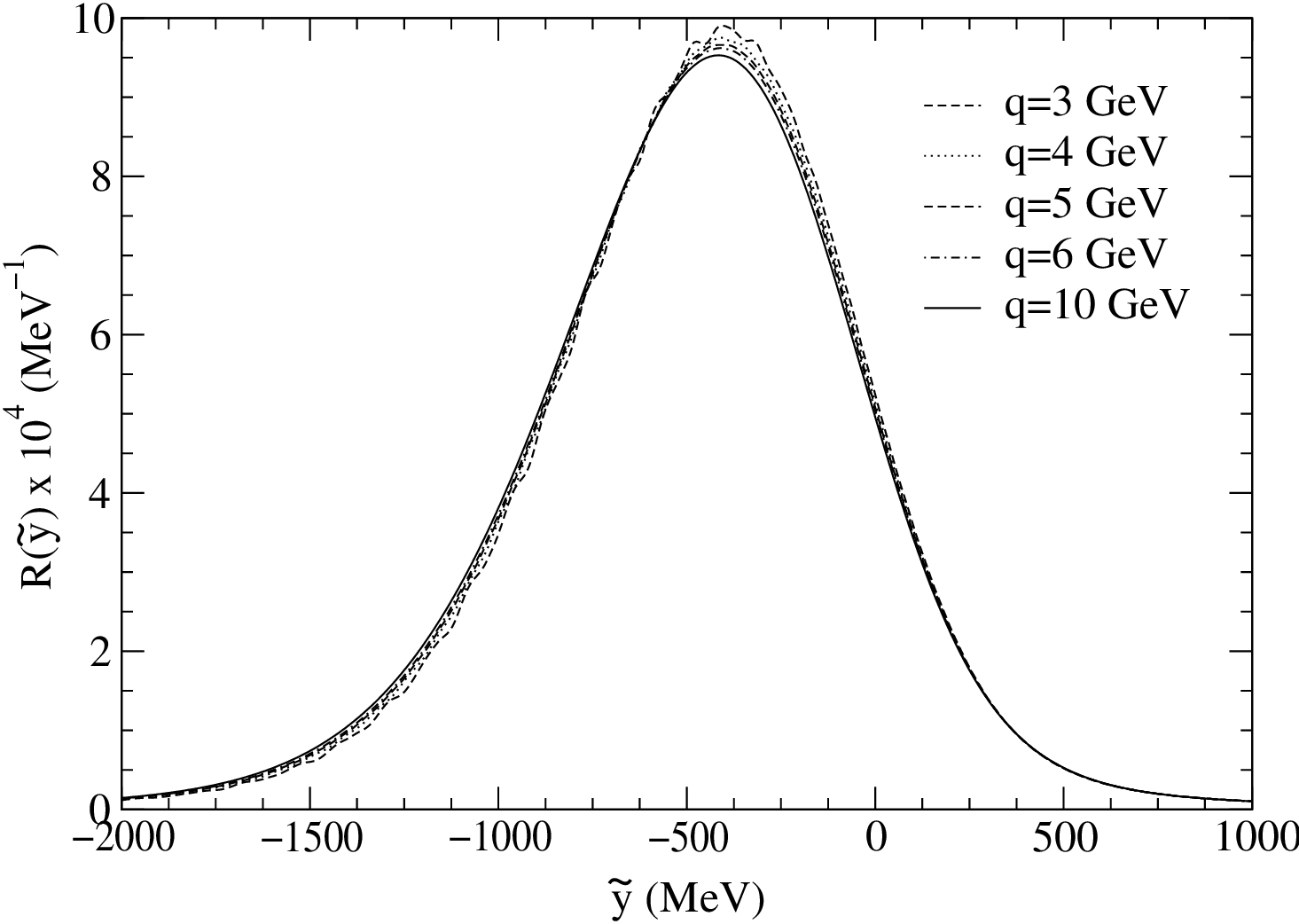}
\nopagebreak
\caption{\label{fig:rhq}
The response for values of $\qmag\geqslant 3$ GeV versus the scaling
variable, $\tilde{y}=\nu-\qmag$.}
\end{figure}

Figure \ref{fig:rhq} shows the response calculated for values of
$\qmag\geqslant 3$ GeV as a function of $\tilde{y}$ for $\Gamma_0 
= 100$ MeV.
The scaling behavior is clearly exhibited; at large $\qmag$ the 
$R(\qmag,\nu)$ becomes a function $f(\tilde{y})$ alone.  This 
scaling is equivalent to $\xi$ scaling via (Eq.~\ref{eqn:xi}).

\begin{figure}[t]
\includegraphics[ width=240pt, keepaspectratio, clip, angle=-90 ]{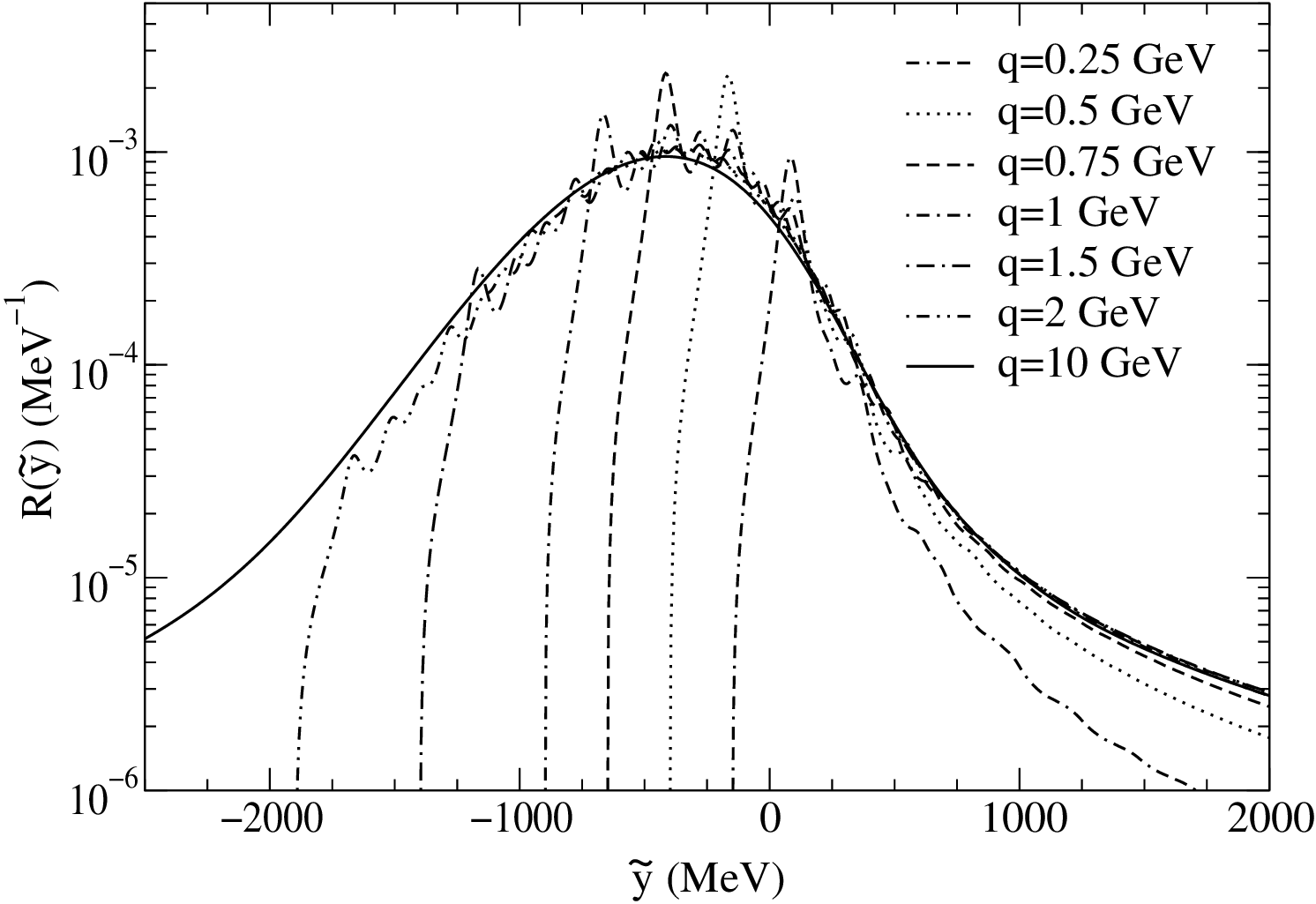}
\nopagebreak
\caption{\label{fig:ras}
The approach to scaling of the response for values of
$\qmag\leqslant 2$ GeV and $\qmag=10$ GeV
versus the scaling variable, $\tilde{y}=\nu-\qmag$.}
\end{figure}

In Fig.\ \ref{fig:ras} we show the response for $\Gamma_0=100$ MeV, at 
various values of $\qmag\leqslant 2$ GeV compared with that for
$\qmag=10$ GeV, to study the approach to scaling.  At small $\qmag$ the 
scattering is dominated by resonances, and the first inelastic
peak is due to the lowest excited state with $n=1$ and $\ell=1$,
335 MeV above the ground state. 
In our toy model, the elastic scattering occurs at $\nu=0$ or
$\tilde{y}=-\qmag$, since our hadron is heavy. 
This elastic scattering
contribution is omitted from Fig.\ \ref{fig:ras}. It has a
strength of $\{0.85,0.53,0.26,0.10,0.010,0.00086\}$
for $\qmag=\{0.25,0.5,0.75,1.0,1.5,2.0\}$ GeV. The energy
dependence of $\Gamma$ [Eq.(\ref{eqn:gam})] implies that the
inelastic response is zero for $\nu < E_t$ or equivalently 
$\tilde{y} < E_t - \qmag $.  

For $\tilde{y} \sim 0$, {\em i.e.} for small $\xi$, 
the response approximately scales at relatively small values of
$\qmag$, comparable to $\sigma^{1/4}$. 
As $\qmag$ increases, the range over which scaling
occurs is extended to more negative 
values of $\tilde{y}$, {\em i.e.} to larger values of $\xi$. 
The contribution of each resonance shifts to lower $\tilde{y}$ 
and decreases in magnitude following the $R(\qmag \rightarrow \infty, 
\tilde{y})$. This behavior is
seen in the experimental data on the proton and deuteron 
\cite{Keppel} and interpreted as evidence for quark-hadron duality. 
Thus the toy model seems to describe some of the observed properties 
of the DIS response of nucleons. 
It exhibits $\tilde{y}$ or equivalently $\xi$ scaling 
at large $|\bvec{q}|$ as observed \cite{BPS00}, 
and an approach to $\xi$ scaling 
similar to that seen in recent experiments.

The $R(\qmag,\nu)$, and therefore the $f(\tilde{y})$ 
extend into the timelike ($\tilde{y} > 0$) region. 
The sum-rule given by Eq.(\ref{eqn:sumr}), 
counts the number of particles in the target. It is 
necessary to integrate over the timelike region to fulfill
this sum rule. In the limit $\Gamma_0 =0$ about 9.6\% of the 
sum is in that region independent of $\st$. It increases to
13.7\% for $\Gamma_0 = 100$ MeV and $\st=1$ GeV/fm. The response
expressed as $R(Q^2,\xi)$ also scales at large $Q^2$ where
$|\bvec{q}|$ is necessarily large. It becomes a function of
$\xi$ alone. However, the integral:
\beq
\int_0^{\infty} R(Q^2 \rightarrow \infty,\xi)d\xi = \int_0^{|\bvec{q}|} 
R(\qmag \rightarrow \infty, \nu) d\nu \leq 0.904,  
\eeq
because the contribution of the timelike region is omitted. The maximum
value, 0.904 of the integral is obtained for $\Gamma_0=0$.
Here we have defined $\xi = \qmag - \nu $ without the conventional 
$1/M$ scale [Eq.(\ref{eqn:xi})].  

The baryon number, or Gross-Llewellyn Smith (GLS) \cite{GLS,ESW}
sum, $S_{GLS}$ is obtained by integrating the neutrino scattering
responses over the range $0<x<1$. In absence of any timelike
response, in the limit $Q^2 \rightarrow \infty$, $S_{GLS}=3$, the
number of quarks minus the number antiquarks in a nucleon. The GLS
sum has gluon radiative corrections that have been calculated at
next-to next-to leading order ({\em i.e.}\ up to
$\mathcal{O}(\alpha_s^3)$) using pQCD \cite{Larin91}. The
experimental values of $S_{GLS}$ obtained from revised CCFR
\cite{CCFR98} $xF_3$ are $2.55 \pm 0.06(\mbox{stat})
\pm 0.10(\mbox{syst})$ and $2.80 \pm 0.13(\mbox{stat})
\pm 0.18(\mbox{syst})$ at $Q^2=3.16$ and 12.59 GeV$^2$,
respectively. They have been used to determine $\alpha_s(Q^2)$
assuming that the difference between 3 and the above experimental
values is entirely due to gluon radiative effects.

Recent measurements of the Drell-Yan cross section ratios,
$\sigma(p+d)/\sigma(p+p)$ from E866 at Fermilab
\cite{Hawker98,JCPeng98} determine the ratio of
$\bar{d}(x)/\bar{u}(x)$ and, in turn, the integral,
\beq
\label{eqn:gsr-sea}
\int_0^1 dx [\bar{u}(x)-\bar{d}(x)] = -0.118\pm0.012
\eeq
to a high level of precision. The valence isospin, or Gottfried, sum
\beq
\label{eqn:gsr}
S_G = \int_0^1 \frac{dx}{x}\left[ F^p_2(x) - F^n_2(x) \right] =
\frac{1}{3} + \frac{2}{3} \int_0^1 dx \left[ \bar{u}(x) - \bar{d}(x) \right]
\eeq
has been measured by the NM Collaboration \cite{Arneodo95}.
The analysis of their complete data set gives
$S_G=0.216\pm0.027$ \cite{Eisele95,Hinch96}
essentially independent of $Q^2$ on the interval
$0.5<Q^2<10$ GeV$^2$. Using these measurements we find a
contribution due to valence quarks of,
\beq
\label{gsr-val}
S_G-\frac{2}{3} \int_0^1 dx \left[ \bar{u}(x) - \bar{d}(x) \right]
= 0.294 \pm 0.030,
\eeq
about 10\% below the expected result of $\frac{1}{3}$,
with comparable error.

The present work suggests that these sums over the spacelike region
could be smaller than the theoretical expectation due
to some of the response being shifted into the timelike region.
Such a shift is due to the bound nature of the quarks in the hadron
and is therefore a nonperturbative effect.

The prevalent models of valence quark structure functions \cite{ESW}
predict a large valence quark response in the spacelike region
adjacent to the photon line which diverges as $\sim x^{-0.5}$.
Even though the $x<0.1$ region is 
dominated by sea quarks, contributions from this region make up a large 
part of the experimental sums; about 50\% of $S_{GLS}$ \cite{CCFR93}
and about 30\% of $S_G$ \cite{NMC94}. Response having $\xi <0.1$
is within $<0.1M\sim 94$ MeV of the photon line. It is possible
that there is some response beyond the photon line, moved into the
timelike region due to nonperturbative effects associated with
binding.

It is known from $e^+e^-$ annihilation experiments that the vacuum 
has timelike response, and the nucleon presumably couples to it.
In the toy model we have ignored this problem in order to focus on
the possibility that the nucleon can have timelike response due to 
the bound quarks in the nucleon being off the mass-shell. 

\acknowledgments

The authors thank Omar Benhar and Ingo Sick for many discussions,
and Jen-Chieh Peng for providing the information on Gottfried sum. 
This work has been partly supported by the US National Science 
Foundation via grant PHY 98-00978.
 
\bibliography{rrf}

\end{document}